\documentclass[preprint]{aastex62}
\received{}
\revised{}
\accepted{}
\submitjournal{AJ}
\shorttitle{Cluster Photometry in M31}
\shortauthors{Leahy et al.}
\begin{document}
\title{AstroSat/UVIT Cluster Photometry in the Northern Disk of M31}

\correspondingauthor{Denis Leahy}
\email{leahy@ucalgary.ca}

\author{Denis Leahy}
\affiliation{Department of Physics and Astronomy, University of Calgary, Calgary, AB T2N 1N4, Canada}

\author{Megan Buick}
\affiliation{Department of Physics and Astronomy, University of Calgary, Calgary, AB T2N 1N4, Canada}

\author{Cameron Leahy}
\affiliation{Department of Physics and Astronomy, University of Calgary, Calgary, AB T2N 1N4, Canada}

\begin{abstract}
 The Andromeda Galaxy (M31) is an object of ongoing study with the Ultraviolet Imaging Telescope (UVIT) on AstroSat. 
 UVIT FUV and NUV photometry is carried out here for a set of 239 clusters in the NE disk and bulge of M31
which overlap with the HST/PHAT survey. 
Padova stellar models were applied to derive ages, masses, metallicities and extinctions 
for 170 clusters.
The ages show a narrow peak at $\sim$4 Myr and  a broad peak around 100 Myr.
log(Z/Z$_{\odot}$) values are mostly between $-0.3$ and $+0.3$.
The 7 clusters in the bulge are low metallicity and high mass.
Most clusters are in the spiral arms and have metallicities in the range noted above. 
The youngest clusters are mostly high metallicity and are concentrated along the brightest parts of the spiral arms.
The UVIT FUV and NUV data are sensitive to young stars and detect a new metal rich peak in star formation in the disk at age $\sim4$ Myr. 
\end{abstract}
\keywords{UV astronomy --- galaxies:M31 --- variability}


\section{Introduction}\label{sec:intro}

The Andromeda Galaxy (M31) is our closest neighbouring giant spiral galaxy and has many similarities to our Galaxy. 
It can serve as a template for studies of parts of our Galaxy that are obscured by extinction.
M31 has a well-measured distance  \citep{2005MNRAS.356..979M} of 785 $\pm25$ kpc (3.2\% error).
The small distance uncertainty leads to well measured luminosities for objects in M31.

M31 has been observed extensively in optical bands.
The Hubble Space Telescope has yielded the highest resolution observations.
These include the Pan-chromatic Hubble Andromeda Treasury (PHAT) survey \citep{2014ApJS..215....9W}. 
The GALEX instrument has observed M31 \citep{2005ApJ...619L...1M} in near and far ultraviolet (NUV and FUV) bands. 

The optical studies have resulted in a wealth of information about M31, including its star formation history (SFH)
and metal enrichment history.
\citet{2018MNRAS.478.5379D} (and references therein) study the SFH of the bulge of M31 and compare 
results from color-magnitude diagram (CMD) analysis with spectral energy distribution (SED) analysis.
\citet{2018MNRAS.475.2754H} (and references therein) explore the merger history of M31, using simulations to explain
long-lived SFH of the 10 kpc ring in the disk.
\citet{2017ApJ...846..145W} map the SFH across the northeast section of M31 covered by the PHAT survey 
($\simeq$1/3 of the whole disk), using a large number ($\sim$3300) of small regions and obtain resolved maps
of SFH and stellar mass. 

More recently, M31 is part of an ongoing survey in NUV and FUV with the UltraViolet Imaging 
Telescope (UVIT) on AstroSat  \citep{2020ApJS..247...47L}. 
The instruments onboard AstroSat include the following: the UltraViolet Imaging Telescope (UVIT) covers the FUV 
and NUV while the Soft X-ray Telescope (SXT), Large Area Proportional Counters (LAXPC), 
Cadmium-Zinc-Telluride Imager (CZTI) and Scanning Sky Monitor (SSM) instruments cover soft through hard X-rays \citep{2014SPIE.9144E..1SS}.

UVIT has high spatial resolution ($\simeq$1 arcsec) which allows individual stellar clusters and stars in M31 to be identified.
 \cite{2021IJAA...11..151L} presents an analysis of the structure and stellar populations of the bulge. 
\cite{Leahy2021} identifies FUV variable sources in the central 28 arcminutes of M31.
\citet{2020ApJS..247...47L} publishes the UVIT point source catalog for M31.
UVIT sources which match with Chandra sources in M31 are analysed by \citet{2020ApJS..250...23L}, and
 initial results of matching UVIT sources with PHAT sources are given by \citet{2021JApA...42...84L}.
\citet{2018AJ....156..269L} analyses UV bright stars in the bulge, showing the existence of young stars in the bulge.

In this work, we analyse the stellar clusters in M31 from the PHAT Stellar Cluster Survey \citep{Johnson} using combined UVIT and HST data. 
We utilize the M31 UVIT observations to obtain FUV and NUV photometry for 239 clusters, and 
provide the Table of UVIT photometry in  Section~\ref{sec:photometry}. 
We fit the combined UVIT and HST photometry using stellar cluster models and give the cluster model fits in Section~\ref{sec:fitting}. 
The model fits are presented and interpreted in Section~\ref{sec:results} and conclusions are summarized in
Section~\ref{sec:conclude}.

\section{Observations and Data Analysis}\label{sec:obs}

UVIT observed M31 galaxy in nineteen 28 arcminute diameter fields, covering NUV to FUV wavelengths \citep{2020ApJS..247...47L}.
 \cite{2017AJ....154..128T} and \cite{2020AJ....159..158T} give a detailed description of the UVIT instrument and its filters,
 which are labelled by their central wavelength in nm. Of the nineteen fields, four fields overlap the region surveyed in PHAT and have measurements in both UVIT NUV filters \citep{Johnson}.

The data processing is carried out using the astrometry corrections given in \citet{2020PASP..132e4503P} and 
data processing and calibration methods given in \citet{2020ApJS..247...47L}.

\subsection{Cluster Selection \& Photometry}\label{sec:photometry}

The UVIT M31 survey (Table 1 and Fig. 2 of \citealt{2020ApJS..247...47L}) covers the PHAT region. 
In the overlap region, UVIT Fields 1, 2, 7, and 13 have both FUV and NUV observations, so we chose those four
fields for our analysis. 
We identified the clusters in \cite{Johnson} that are located in these four fields, then 
carried out photometry at the cluster positions in the UVIT data using CCDLAB (\citealt{2017PASP..129k5002P}, 
\citealt{2020PASP..132e4503P}, \citealt{2021JApA...42...30P}). 

We used a fitting box size of 9 pixels ($\sim 3.7\arcsec$) to correspond with previously calculated photometric conversions 
from \cite{Leahy2021}. 
As a result, we restricted our analysis to sources with an angular diameter less than $1\arcsec$ in \cite{Johnson} to ensure accurate photometric measurements. 
This process yielded 183 clusters in Field 1, 335 clusters in Field 2, 268 clusters in Field 7, and 184 clusters in Field 13. 
For the UVIT photometry, if the magnitude calculated by CCDLAB was found at a position more than $1\arcsec$ from the initial position in \cite{Johnson}, we consider that to be a non-detection and list it as a missing value.

Of the 970 clusters, we selected 281 cluster measurements for which we have a magnitude or upper limit for every filter 
from \cite{Johnson} and a magnitude brighter than 23 for UVIT filters F148W, F172M, N219M and 
N279N\footnote{This corresponds approximately to the detection limit of the UVIT Fields.}. 
Most Field 1 clusters also have a F169M magnitude. 
Of the cluster measurements, 84 of them are pairs of measurements from two adjacent fields. 
Thus, we combine these flux measurements into 42 unique cluster measurements weighted by their exposure times. 
This gives a total of 239 clusters. Of these: 238 have F148W, N219M and N279N magnitudes; 
239 have F172M magnitudes; and 70 have F169M magnitudes. 
The photometry measurements are given in Table~\ref{tab:Photo} (the full Table is given online).  

The ranges of UVIT AB magnitudes for the 239 clusters is: 17.26 to 23, mean 20.66 for F148W; 16.83 to 23, 
mean 20.63 for F172M; 17.26 to 23, mean 20.59 for N219M; 16.64 to 23, mean 20.10 for N279N; 
and  17.76 to 23, mean 20.66 for F169M. 
Because the errors in the N279N measurements are significantly larger than the PHAT F275W measurements 
and the magnitudes are consistent, we omit the N279N values from Table~\ref{tab:Photo}.

Systematic errors in the photometry include $\sim$1\% calibration uncertainty \citep{2017AJ....154..128T}, and uncertainty in the correction factor \citep{Leahy2021} of several percent  to account for
overlap in the extended wings of the UVIT point-spread-function \citep{2017AJ....154..128T}.
There are also systematic errors in the cluster models of a few percent (e.g. see discussion in \citealt{2017ApJ...846..145W}).
To account for these systematic errors in the photometry we tried two cases: i) add 0.05 magnitude error and ii) add 0.1 magntude error in quadrature to the statistical error for the magnitudes. 

 \begin{figure}[htbp]
    \centering
    \epsscale{1.2}
    \plotone{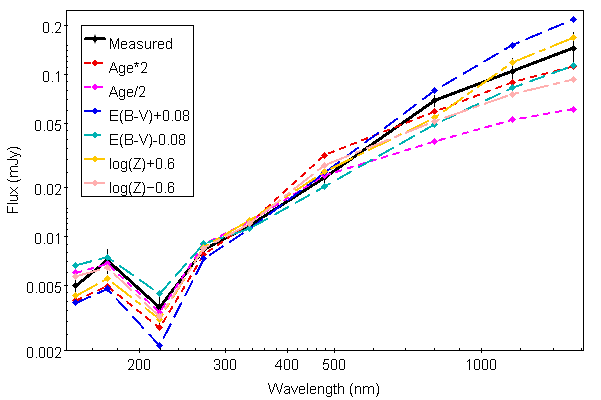}
   \figurenum{1}
    \label{fig:sample}
   \caption{The cluster model fit to UVIT and PHAT magnitudes for Source 1596  in the catalog of \cite{Johnson} 
   (number 84 in our catalog). The black dots with error bars are the data and the black line connects 
   the best-fit cluster model photometry points. The other six lines are for cluster models with the best-fit parameters fixed except for the labelled parameter: 
   i.e. age increased (red) or decreased (magenta) by a factor of 2; 
   log(Z/Z$_{\odot}$) increased (yellow) or decreased (pink) by 0.6; 
   or E(B-V) increased (blue) or decreased (cyan) by 0.08. The 6 model curves with altered parameters were normalized to the data at 336 nm in order to illustrate changes in shape.}
\end{figure}

\subsection{Cluster Model Fits}\label{sec:fitting}

To obtain cluster properties (mass, age, metallicity and extinction), 
we created a Python program to model the UVIT and PHAT magnitudes with published stellar cluster models.
We chose to use the  Padova stellar models with Kroupa initial mass function  \citep{2001MNRAS.322..231K}, 
as did the work of \cite{Johnson2016}.

The models were calculated using the CMD 3.4 online tool at http://stev.oapd.inaf.it/cgi-bin/cmd.
\cite{2022AJ....163..138L} gives a more complete description of the program.
The program reads in a large grid of Padova stellar models with different metallicities 
(log(Z/Z$_{\odot}$)= -2.616 to 0.303) and age ($3.98\times10^{6}$ yr to $1.5\times10^{10}$ yr), then carries out a 
2-dimensional interpolation to obtain the multi-band fluxes for any age and metallicity within those ranges. 
Extinction is calculated using the extinction law of \cite{2007ApJ...663..320F}.
The program finds the best-fit stellar cluster parameters (age,  log(Z/Z$_{\odot}$), mass and
extinction E(B-V)) by $\chi^2$ minimization. 
For each cluster we calculated fits for a grid of parameters around the best-fit parameters 
to determine the 1$\sigma$ parameter uncertainties.

For the 239 clusters, we fit the combined HST F275W, F336W, F475W, F814W, F110W and F160W filter 
photometry from \cite{Johnson} with the UVIT filter F148W, F172M, F169M, and N219M photometry.
We chose not to fit the N279N filter because it covers essentially the same waveband as the F275W filter and has significantly larger uncertainties. 
We added a systematic error to the photometry as noted above.
The fits with 0.05 magnitude systematic errors gave higher $\chi^2$ than expected for the fits and unrealistically small parameter errors (e.g. compared to those from \citealt{Johnson2016}). 
The second case using 0.1 magnitude gave reasonable $\chi^2$ and parameter errors, so we adopt that case.

For some of the clusters we were not able to obtain a reasonable cluster model fit to the data. 
These fits were identified manually and consisted of two groups: either the cluster models were not able to fit the photometry data (poor fits), or
the photometry data was poor by either missing several bands or with large photometry large errors (poor data). 
Possible reasons for the first group (poor fits) are:  
the photometry might be contaminated by non-cluster members in the same line-of-sight as the cluster; 
the UV light could be dominated by rare massive stars so that stochastic variation in the number of massive stars causes large stochastic variation in the UV photometry; 
interacting binaries which are not in the cluster models could contribute to the UV light; 
the standard extinction law \citep{2007ApJ...663..320F} which we used might not be applicable to some clusters.

The net result was that we obtained model fits for 170 clusters with UVIT photometry (labelled UVIT-PHAT clusters).
Using the fitting program, we obtained best-fit values of log(Z/Z$_{\odot}$), log(age), log(M/M$_{\odot}$), and E(B-V) for these clusters. 
Log(Z/Z$_{\odot}$) and log(age) were allowed to vary in the above specified ranges. 
E(B-V) was allowed to vary between 0 and 1 and the normalization (log(M/M$_{\odot}$)) was allowed to vary freely.
The parameters errors were determined by calculating $\chi^2$ values for a finely spaced grid of parameters around the best-fit values of  log(age), log(Z/Z$_{\odot}$) and E(B-V).
For each set of these 3 parameters, the normalization (log(M/M$_{\odot}$)) was minimized.

Figure \ref{fig:sample} shows an example best-fit for one of our sources. 
To illustrate the changes in photometry which result from variation of single cluster parameters,
Figure \ref{fig:sample} shows six additional model curves. 
We renormalize each curve to match the data at 336 nm in order to emphasize the difference in shape of the different curves. 
Increasing age by a factor of two results in a redder spectrum (faint in FUV and NUV, bright in NIR)
and decreasing age by factor two has the opposite effect. 
Changing extinction, E(B-V) by $\pm$0.08 has a similar effect: a redder spectrum for higher extinction and bluer spectrum for lower extinction. 
Extinction and age are only partially degenerate because they have quite different effects for the intermediate wavelengths (between 300 and 1000 nm) and the depth of the 220 nm dust absorption peak is dependent on extinction but not age. 
Changing metallicity, Z, has a smaller but similar effect to age in changing the cluster model photometry.
The effect of changing mass (normalization of the photometry) is not shown because it does not change the shape of the spectrum.

\begin{figure}[htbp]
    \centering
    \epsscale{1.2}
    \plotone{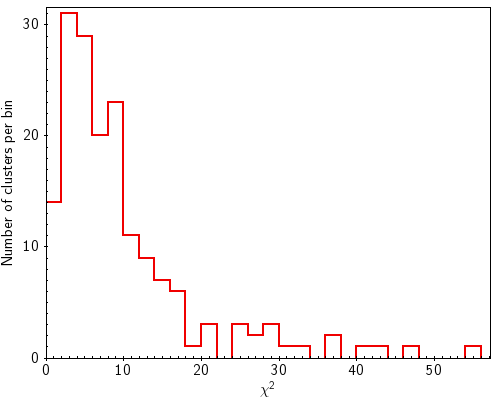}
    \figurenum{2}
    \label{fig:chi2}
   \caption{Histograms of best-fit $\chi^2$ for the UVIT-PHAT clusters.
   }
\end{figure}

\begin{figure}[htbp]
    \centering
    \epsscale{1.2}
    \plotone{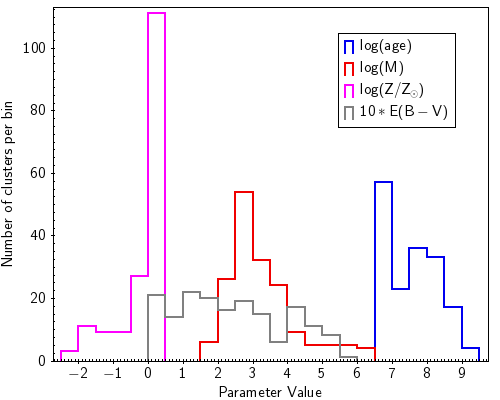}
    \figurenum{3}
    \label{fig:hist}
   \caption{Histograms of best-fit parameters  for the UVIT-PHAT clusters. 
    E(B-V) is multiplied by 10 so that it is on a similar x-axis scale to 
   the other parameters.}
\end{figure}

The parameters from the cluster model fitting are given in Table 2.
The distribution of best-fit $\chi^2$ values  is shown in Figure~\ref{fig:chi2}.
Table~\ref{tab:FitStats} shows the statistics of the  $\chi^2$ values, the fit parameters and their errors.

\section{Results and Discussion}\label{sec:results}

\begin{figure}[htbp]
    \centering
    \epsscale{2.3}
    \plottwo{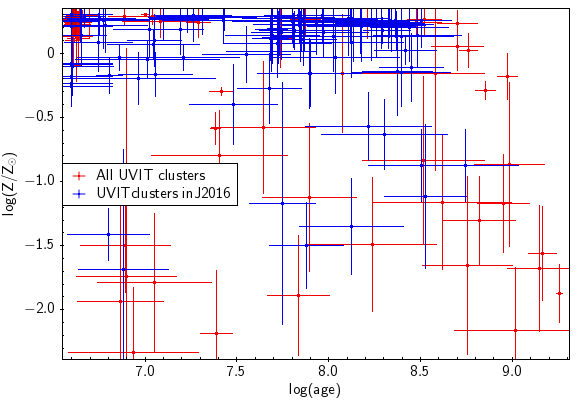}{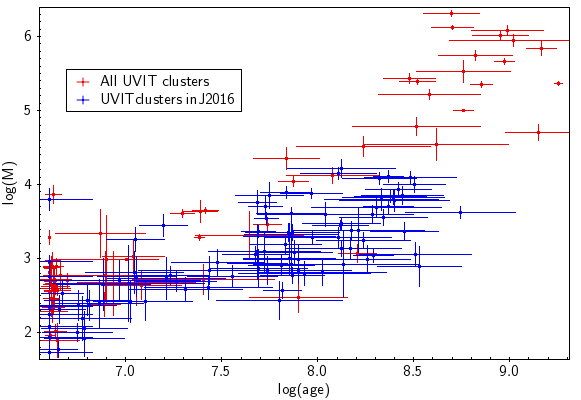}
    \figurenum{4}
    \label{fig:XY}
   \caption{Top panel: Metallicity vs. age with errors, illustrating that the most clusters have 
   log(Z/Z$_{\odot}$) between -0.3 and +0.3.  
   Bottom panel: Mass vs. age with errors, showing a monotonically increasing average mass with age.
   }
\end{figure}

\subsection{Cluster Masses, Ages, Metallicities and Extinctions}\label{sec:distns}

The histogram of best-fit cluster parameters is shown in Figure~\ref{fig:hist}.
Fit masses range from $\sim$60 to $2\times10^6$ M$_{\odot}$, with mean error of a factor of 1.4
(0.15 error in log(M/ M$_{\odot}$).
The mass distribution has a single broad peak at  $\sim$1000 M$_{\odot}$.
Fit ages range from $4\times10^6$ to $2\times10^9$ yr, with  mean error of a factor of 1.5.  
There are a narrow age peak at $4\times10^6$ and a broad peak around $1\times10^8$ yr. 

\begin{figure*}[htbp]
    \centering
    \epsscale{1.2}
    \plotone{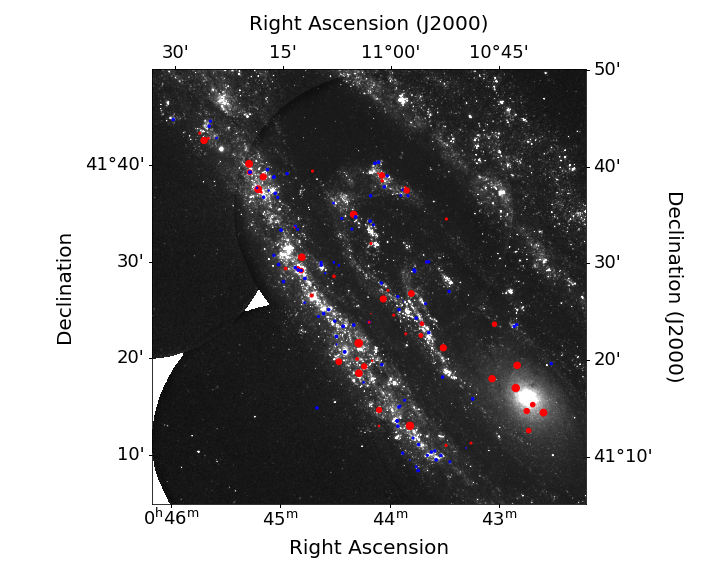}
    \figurenum{5}
    \label{fig:Zposn}
   \caption{Overlay of cluster positions (solid circles) on the UVIT F148W image of the central and northeast parts of M31
   \citep{2020ApJS..247...47L}. 
   Log(Z/Z$_{\odot}$) values are shown by the size of the circles with radius proportional to log(Z/Z$_{\odot}$), 
   with blue indicating positive values and red indicating negative values, and total range -2.3 to +0.3.}
\end{figure*}

Fit metallicities, log(Z/Z$_{\odot}$), range from greatly subsolar (-2.3) to supersolar (+0.3), 
with mean error of 0.24.
There is a peak with log(Z/Z$_{\odot}$) $\simeq$-0.3 to +0.3.
Fit extinctions, E(B-V), have a fairly smooth distribution over the range from 0 to 0.6.
E(B-V) mean fit errors are 0.08.

The metallicities of the clusters are shown plotted against age (top panel of Figure~\ref{fig:XY}).
99 of the UVIT clusters were previously studied by \cite{Johnson2016} using color-magnitude-diagram (CMD) analysis. These are marked in  Figure~\ref{fig:XY} by the blue points.
Most clusters ($\sim80$\%) have log(Z/Z$_{\odot}$) above -0.3.
No clear correlation of metallicity with age is seen, except that the oldest (log(age)$>$8.7) are all 
metal-poor.
The masses vs. ages are shown in the bottom panel of Figure~\ref{fig:XY}.
There is a  clear correlation of mass with age: older clusters are more massive.

\subsection{Spatial Distribution of Cluster Properties}\label{sec:spacedist}

\begin{figure*}[htbp]
    \centering
    \epsscale{1.2}
    \plotone{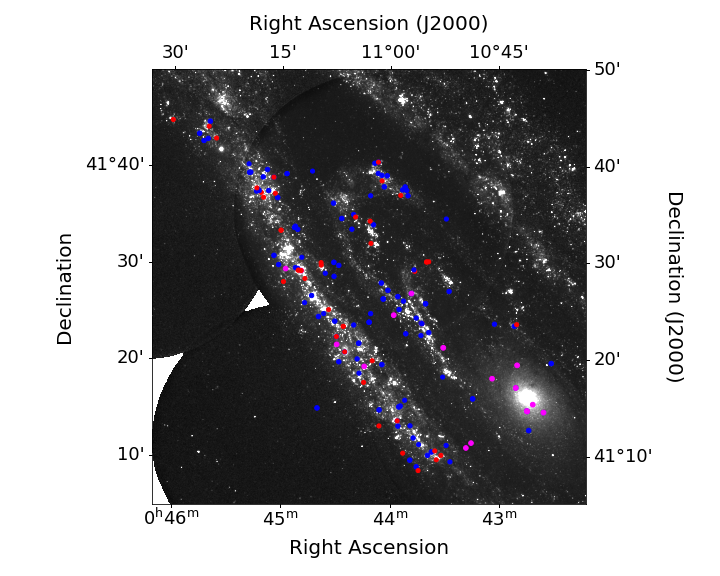}
    \figurenum{6}
    \label{fig:ageposn}
   \caption{Overlay of cluster positions (solid circles) on the UVIT F148W image of the central and northeast parts of M31. 
   Log(age/yr) values are shown by the size of the circles with radius proportional to log(age/yr) and total range 6.6 to 9.7. Clusters with log(age/yr)$<$6.75 are in red; 6.75$<$log(age/yr)$<$8.55 are in blue; and log(age/yr)$>$8.55 are
   in magenta.}
\end{figure*}

The locations of the UVIT-PHAT clusters are plotted in Figure~\ref{fig:Zposn} on top of the UVIT 
F148W mosaic image of M31 \citep{2020ApJS..247...47L}.
The log(Z/Z$_{\odot}$) values are shown by the radii of the circles at the cluster positions. 
The majority of the clusters are located within the UV-bright spiral arms, with a small subset of $\sim$7 clusters around
the central bulge of M31 and $\sim$10 clusters located between the arms. 
The clusters around the bulge are all low metallicity, whereas the arm and interarm clusters show a wide range of metallicity. 
Along the main spiral arm running diagonally across the image from NE to SW the metallicity of clusters are mostly positive but about 20\% have negative values. 
The inner arm has a similar metallicity pattern. 

Figure~\ref{fig:ageposn} shows the ages of the clusters on the M31 F148W image. 
The youngest clusters are located along the central bright ridges of the spiral arms. This can be 
seen, e.g. for the clusters near R.A 10$^\circ$55$\arcmin$, Dec. 41$^\circ$12$\arcmin$ or clusters near
 R.A 11$^\circ$18$\arcmin$, Dec. 41$^\circ$37$\arcmin$.
Both groups of clusters are located on the main 
spiral arm running diagonally from southwest to northeast across the image. 
6 of the 7 clusters around the bulge are old (age$>3\times10^{8}$ yr, in magenta), with the other old clusters not clearly correlated with spiral arm positions. 
The intermediate age clusters (in blue) are mostly near the spiral arms, with $\sim10$ to 20\% in the interarm or near bulge regions.

 The masses of the clusters (not shown) look very similar to the ages on the M31 F148W image. 
The low-mass clusters, are located along the spiral arms at the same positions
as the low-age clusters. 
This is consistent with the correlation of mass and age shown in Figure~\ref{fig:XY}.
The 7 clusters around the bulge are all massive (M$>10^{5}$ M$_{\odot}$) as well as low metallicity (Fig.~\ref{fig:Zposn}).  A few other massive and low metallicity clusters are found in interarm regions.

\begin{figure}[htbp]
    \centering
  \epsscale{1.2}
    \plotone{{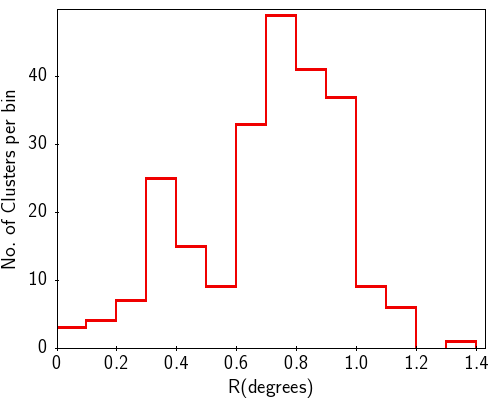}}
    \figurenum{7}
    \label{fig:Nbin}
   \caption{Number of clusters in bins (0.1$^\circ$ width) of distance, $R$, in the disk plane from the centre of M31. }
\end{figure}

The cluster extinctions, E(B-V), (not shown) have no correlation with the positions of the spiral arms or the bulge. 
The clusters near the bulge or the spiral arms can have high or low extinction 
with high extinction clusters and low extinction clusters often near eachother. 
This is consistent with extinction caused by line-of-sight depth into M31.

\begin{figure*}[htbp]
    \centering
    \epsscale{1.2}
    \gridline{\fig{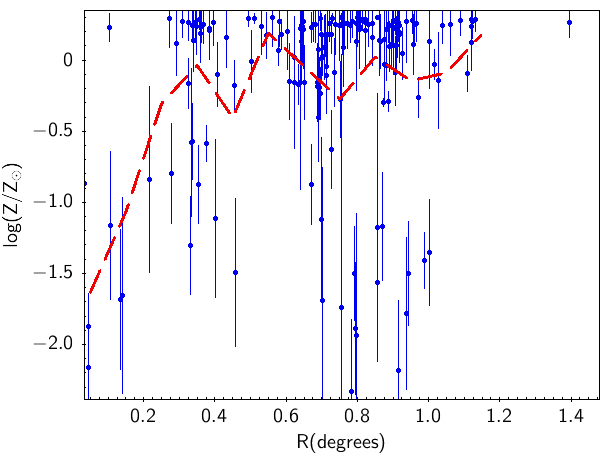}{0.5\textwidth}{(a)}
             \fig{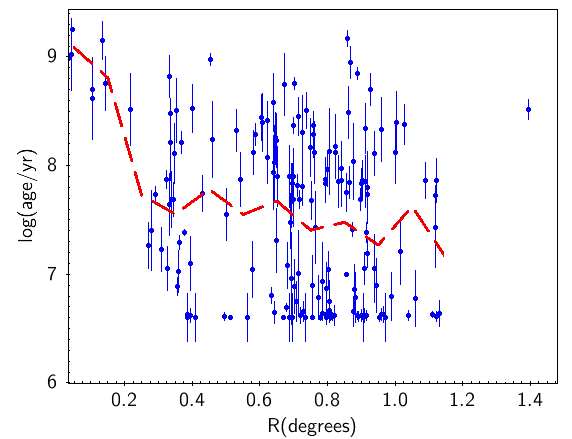}{0.5\textwidth}{(b)}}
    \gridline{\fig{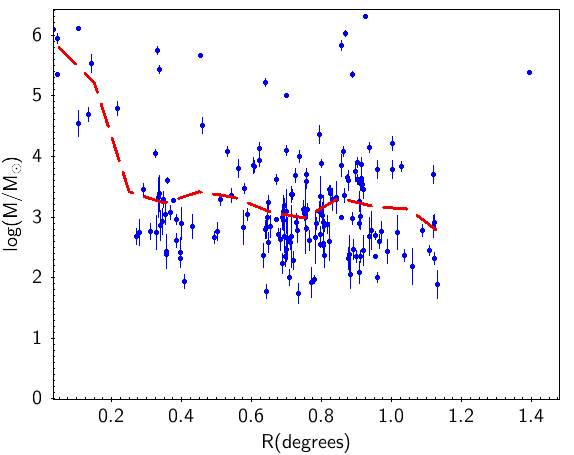}{0.5\textwidth}{(c)}
          \fig{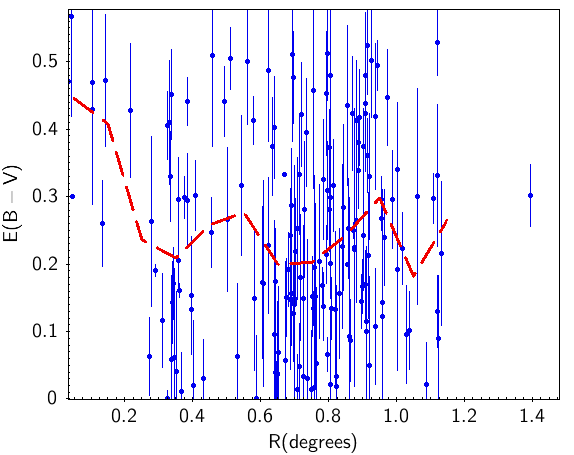}{0.5\textwidth}{(d)}}
    \figurenum{8}
    \label{fig:radialplot}
   \caption{Panels (a), (b), (c) and (d): values of log(Z/Z$_{\odot}$), log(Age/yr), log(M/M$_{\odot}$), 
   and E(B-V) respectively for each cluster (blue dots with error bars) and average values per bin (red dots with dashed line).}
\end{figure*}

To test for variations of cluster properties from the centre of M31, assuming the clusters are concentrated in the disk
of M31, we correct for the inclination of M31, which is $\simeq77^\circ$. 
The deprojected distance from the centre in the disk plane, $R$ in degrees, is calculated for each cluster.
At the distance of M31,  0.1$^\circ$ corresponds to a distance of 1.37 kpc.

The number of clusters vs. $R$ is shown in Figure~\ref{fig:Nbin}.
The two peaks in the number of clusters are from the clusters associated with the inner arm at a 
distance of 4.8 kpc and the clusters associated with the main spiral arm at a distance of 10 kpc from the centre of M31.
The main spiral arm has $\sim130$ UVIT-PHAT clusters in it, and the inner arm has $\sim40$ clusters.  
The peaks are broader than the width of each spiral arm because R of the center of the spiral arms 
changes with position angle.

The metallicities of the clusters 
with model fits are shown in  Figure~\ref{fig:radialplot}.
Most disk clusters (outside of 0.3$^\circ$) have log(Z/Z$_{\odot}$)  between -0.3 to +0.3.
1.2$^\circ$)
In the bulge region (inside 0.3$^\circ$) most are metal poor\footnote{The cluster with high metallicity 
at $\sim$0.1$^\circ$ appears inside a few of the bulge clusters because we applied the inclination 
correction to the bulge clusters as well as disk clusters.}

Panel (b) 
shows the ages of the clusters. Inside 0.3$^\circ$ (4 kpc), there are only old clusters. 
In the dense regions of the inner and main spiral arms, there are old, intermediate and young clusters. 
The few old clusters might be bulge clusters in projection on the disk.
The young clusters ($\sim10^7$ yr)  are smaller in number (by factor $\sim$2)
than the intermediate age clusters ($\sim10^8$ yr).

Panel (c) 
shows masses of the clusters. 
Inside 0.3$^\circ$ (4 kpc), there are only massive ($>2\times10^4$ M$_{\odot}$) clusters.
In the regions of the inner and main spiral arms, there are predominantly low mass clusters. 
The few high mass clusters ($>10^5$ M$_{\odot}$) outside 0.3$^\circ$ might be bulge clusters in projection on the disk.

Extinctions of the clusters are shown in panel (d). 
 Inside 0.3$^\circ$, E(B-V) values are higher (0.3 to 0.7), whereas from 0.3 to 1.2$^\circ$
 E(B-V) is nearly uniformly spread between 0 and 0.5. 
This is consistent with being caused by line-of-sight depth differences to different clusters.

The mean parameter values per bin are shown in Figure~\ref{fig:radialplot} by the points connected by dashed
lines. These show the mean cluster metallicities in the bulge are lower than for the spiral arm regions.
The mean ages and masses are much smaller for the spiral arm regions than the bulge (by factors of $\sim100$
for both). The mean extinction for the bulge is higher than for the spiral arms.

Comparing the inner and main spiral arms, the clusters have similar metallicity distributions. 
However the main arm has significantly higher proportion of young ($\sim 10^7$ yr) clusters than the inner arm.
The mass distributions are not significantly different.
The lower mean mass for the main arm is caused by the higher numbers of $10^2$ to $10^4$M$_{\odot}$ clusters.
The mean extinction values for the main arm and the inner arm are the same.
 
\begin{figure}[htbp]
    \centering
    \epsscale{2.2}
\plottwo{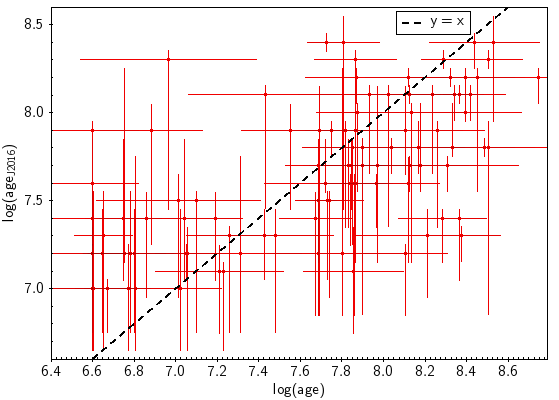}{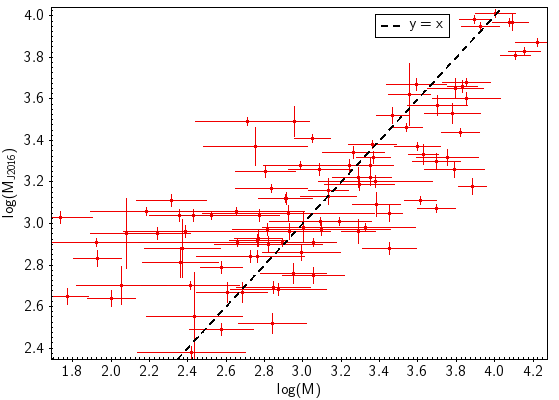}
    \figurenum{9}
    \label{fig:logMvlogA}
   \caption{The UVIT-PHAT results (x-axes) for  log(age)  (top panel) and
   log(M) (bottom panel) compared to the results  from \cite{Johnson2016} (y-axes) for the 99 clusters in both samples.
   Errors are $\pm$1$\sigma$ for both cases. 
   }
\end{figure}

\subsection{Comparison with previous results}\label{sec:compare}

\subsubsection{Comparison with PHAT\citep{Johnson2016} cluster analysis}

\cite{Johnson2016} analyzed the subset of 1249 of the 2753 clusters from \cite{Johnson} with ages  $<300$ Myr 
to study the cluster formation efficiency ($\Gamma$) in M31. This included photometry of resolved stars in each 
cluster and F475W vs F475W-F814W CMD fitting to determine age, mass and extinction ($A_V$). 
A small variation in metallicity was included ($-0.2<$log(Z/Z$_{\odot}<0.1$). 
Their catalog of clusters gave ages and masses but not extinction or metallicity.
They obtained maps of star formation rate (SFR) surface density for two age bins, 10-100 Myr and 100-300 Myr, 
and obtained $\Gamma$ in seven spatial bins for the two age bins.

The analysis done here is significantly different. 
Whole cluster photometry (UVIT and PHAT \citealt{Johnson}) is obtained for 9 wavebands from FUV to NIR
and used to obtain age, mass, metallicity and extinction. 
Larger ranges of age and metallicity is used compared to \cite{Johnson2016}. 
Both studies use the Padova cluster models. 

A subset of 99 clusters are in both studies.
Figure \ref{fig:logMvlogA} shows our results (x-axis) and the \cite{Johnson2016} results (y-axis, labelled J2016) for ages and masses.
A wider range of ages and of masses are found in our study. 
The wider range of ages is likely caused by two factors: we include FUV and NUV bands which are  sensitive to age,
and the study of \citet{Johnson2016} had a narrower range in their model age grid ($10^7$ yr to $3\times10^8$ yr).
Their clusters near their lower or upper limits could be fit better by ages below their lower limit or above their upper limit. 
The change in age affects the normalization and thus the mass of each cluster, with generally less mass required
for lower age because the stars are on average brighter.

The FUV and NUV photometry of the current work should be more sensitive to extinction, which peaks near 200 nm, than previous studies.
This is likely a factor in the differences in ages and masses. 
It can be seen from Figure~\ref{fig:logMvlogA} that the UVIT age errors are similar to, and the UVIT mass errors are larger than those of \cite{Johnson2016}. 
Nearly all of the UVIT ages are less than 2 sigma different than the \cite{Johnson2016} values.
For masses, about 75\% are less than 2 sigma different.

\begin{figure}[htbp]
    \centering
    \epsscale{1.1}
    \plotone{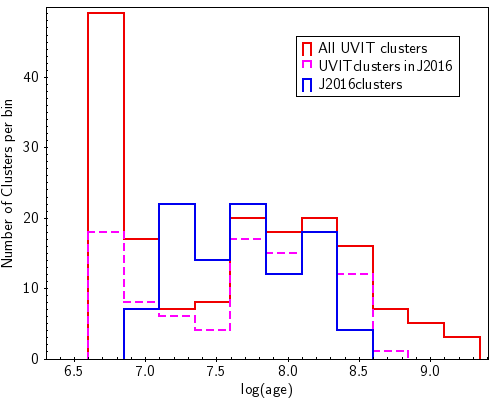}
    \figurenum{10}
    \label{fig:UVTv2017W}
   \caption{
  Distributions of age for the 134 clusters in the UVIT-PHAT sample (this work), with cluster model lower  log(age) limit of 6.6, and the \citet{Johnson2016} sample, with cluster model log(age) limits of 7.0 and 8.5.
   }
\end{figure}

Figure~\ref{fig:UVTv2017W} shows the distribution of ages from our photometry analysis of the UVIT clusters 
(all 170 and the 99 in common with \cite{Johnson2016}) and ages for the 99 clusters with CMD analysis from  \cite{Johnson2016}.
For the 99 clusters, the UVIT/photometry age distribution (dashed magenta line) is similar to the that from the CMD analysis
(blue) but with slightly large age range. The full sample of 170 UVIT clusters (red) shows an extra peak at ages below $10^7$ yr and a tail above $3\times10^8$ yr extending to $2\times10^9$ yr.
Thus the set of 71 UVIT clusters not in the sample of \cite{Johnson2016} contains most of the youngest and the oldest clusters.  

In summary, our method and that of \cite{Johnson2016} each have their own strengths and weaknesses.
Figure~\ref{fig:logMvlogA} shows that the clusters in common have differences in ages and masses which are mostly not significant.
The method of \cite{Johnson2016} works well for clusters diffuse enough to be dominated by resolved stars. 
The UVIT data has sufficiently spatial resolution (1$\arcsec$) to measure whole-cluster photometry, but not individual stars in each cluster.
The advantage of the whole-cluster photometry is that it includes light from the whole cluster, including faint and unresolved stars, 
whereas the advantage of CMD fitting is that it contains information for those stars that are resolved. 
Another advantage of the UVIT photometry for young clusters is that the stars have significant FUV and NUV emission 
which is not analyzed in the F475W-F814W optical wavelength CMDs.

\subsubsection{Comparison with SFH analysis from starlight in the disk and bulge of M31}

The SFH of the disk has been extensively studied  using CMD analysis on large 
numbers of resolved stars from PHAT \citep{2017ApJ...846..145W}. 
The main results from that study include a table of SF rates in 
$\sim$3300 spatial bins for 16 age bins from 0 to 14.1 Gyr, metal enrichment history for 3 radial bins in
the disk, and history of stellar mass build-up in the disk. Parameters derived using four different 
stellar evolution codes, including the Padova models, were compared. 
They find most of M31's disk stars were formed $>$8 Gyr, there was a widespread SF episode of age $\sim2$ Gyr and
the 10 kpc star-forming ring (the main spiral arm in Fig.~\ref{fig:Zposn}) is visible for all ages 
$\lesssim$1 Gyr.
Their smallest age bin is from $4\times10^6$ to $3.2\times10^8$ yr, 
which covers almost the entire age range found here for the UVIT clusters. 
The current study selects stars in clusters rather than fields stars, which make up most of the stellar mass. 
Cluster studies are biased toward the youngest stars that form in any SF episode
because a higher fraction of clusters disperse for higher age  
\citep{2012MNRAS.425..450M}\footnote{Only if SF occurs in short bursts will the clusters have the same ages as the SF episodes.}.

Studies of the SFH of the bulge 
include \cite{2022AJ....163..138L}, \cite{2018MNRAS.478.5379D}, \cite{2018A&A...618A.156S}, and references therein. 
These find two and possibly three episodes of SF in the bulge, with
ages of 10-12 Gyr with log(Z/Z$_{\odot}$)$\simeq$0.3 and $>90$\% of the total stellar mass, $\sim$600 Myr with log(Z/Z$_{\odot}$)$\simeq$0 and $\sim5$\% of the mass, 
and $<100$ Myr \citep{2018MNRAS.478.5379D} or $\sim$25 Myr  \citep{2022AJ....163..138L} 
with log(Z/Z$_{\odot}$)$\sim$-0.7 and a small fraction ($\sim10^{-6}$) of the mass. 
The old broad peak detected here and in \cite{Johnson2016} between $\sim$20 and 300 Myr (Figure~\ref{fig:UVTv2017W})
has age between those of the $\sim$600 Myr SF peak and the youngest peak at $\sim$25 to 100 Myr found by the studies in the bulge above.
This and the different locations means that they are likely not physically related.

The youngest SF peak detected here, with age $\sim 5\times10^6$ yr and mean log(Z/Z$_{\odot}$) of +0.18, 
might be related to the detection of similar aged stars in the bulge,  
by the study of UV-bright stars in the bulge \citep{2018AJ....156..269L}.
This youngest peak was probably not detected by \citet{Johnson2016} because of their lower age cutoff of $10^7$ yr. The  FUV and NUV data from UVIT was also essential to detect these young stars here.

\section{Conclusions}\label{sec:conclude}

In this study, UVIT FUV and NUV photometry was carried out for a set of 239 clusters  
which overlap with the HST/PHAT survey. The area covers the NE disk and bulge of M31. 
The FUV to NIR SEDs were constructed from our UVIT photometry and the PHAT photometry \citep{Johnson} 
and were modelled for 170 clusters using 
our cluster SED fitting program with Padova stellar models to derive age, mass, metallicity and extinction.

The best-fit models for the clusters show a range of masses 
from  $\sim$60 to $2\times10^6$ M$_{\odot}$. 
The clusters have a narrow age peak at $4\times10^6$ and a broad peak around $1\times10^8$ yr. 
Higher age clusters generally have higher masses. 
Metallicities (log(Z/Z$_{\odot}$)) of the clusters are mostly in the range -0.3 to +0.3.
Extinction, E(B-V), covers a broad range from 0 to $\sim$0.5.
Extinction is not correlated with position, consistent with clusters located at different line-of-sight depths independent
of sky position.
The spatial distributions of the UVIT-detected clusters and their parameters 
high metallicity/low mass clusters are more spatially 
correlated with each other and more concentrated towards the brightest parts of the spiral arms,
and the low metallicity/high mass clusters are spread fairly uniformly.

Previous work on stellar ages and metallicities in the disk of M31 \citep{Johnson2016} 
fit resolved HST CMDs of clusters in the PHAT survey.  
99 of those clusters are in our UVIT sample. 
We find a broader range of ages and masses for the clusters, in part because we allow a wider range of ages 
and metallicities. 
However the differences are less than 2 sigma for the majority of clusters.
This is consistent with the studies of \citet{2021ApJ...913...45O} and \citet{2013ApJ...772....8J}   which compare CMD and SED fitting methods.

For field stars in disk of M31,  \citet{2017ApJ...846..145W} 
carried out CMD analysis using PHAT photometry for $\sim$3300 spatial regions.
They find most stars are older than 8 Gyr, widespread SF with age $\sim$2 Gyr, and a weak peak
with age $\sim$0.6 to $1.2~\times10^9$ yr. 
The study here of clusters and the restriction to those detected in FUV and NUV means
the current work detects younger stars
in contrast to their study which detects older SF episodes.

\citet{2022AJ....163..138L} combined UVIT data with optical and IR data for the bulge of M31.
SED analysis gave results consistent with that from HST/PHAT data \citet{2018MNRAS.478.5379D}.
The M31 bulge has a dominant $\sim$10 Gyr metal-rich  population, a younger 
$\sim$600 Myr $\sim$solar-metallicity population, and a small fraction of metal poor 
 stars with age $\sim$25 to 100 Myr.
The current study of UVIT clusters in the disk detects clusters 
with ages in the range of 10-100 Myr  and a younger set of clusters with ages $\sim$4 Myr.
The first set has ages similar to the youngest population in the  bulge but has a different
metallicity distribution. The second UVIT set is a newly detected population, which should be followed up in more detail by new observations and analysis.

\acknowledgments

This work is undertaken with the financial support of the Canadian Space Agency and of the Natural
Sciences and Engineering Research Council of Canada. This publication uses data from the AstroSat 
mission of the Indian Space Research Institute (ISRO), archived at the Indian Space
Science Data Center (ISSDC).

\begin{longrotatetable}
\begin{deluxetable*}{ccccccccccccc}
\tablecaption{UVIT photometry$^a$ of M31 PHAT stellar clusters.\label{tab:Photo}}
\tablewidth{700pt}
\tabletypesize{\scriptsize}
    \tablehead{\colhead{Source $\rm{ID^b}$} & \colhead{UVIT} & \colhead{UVIT} & \colhead{F148W} & \colhead{F148W} & \colhead{F172M} & \colhead{F172M} & \colhead{N219M} & \colhead{N219M} & \colhead{F169M} & \colhead{F169M} & \colhead{Separation} & \colhead{Region}
    \\ \colhead{} & \colhead{$\rm{RA^{c}}$} & \colhead{$\rm{DEC^{c}}$} & \colhead{AB} & \colhead{err} & \colhead{AB} & \colhead{err} & \colhead{AB} & \colhead{err} & \colhead{AB} & \colhead{err} & \colhead{} & \colhead{}
    }
    \startdata
    16  & 10.93972 & 41.14929 & 22.51 & 0.10 & 21.93 & 0.14 & 22.13 & 0.21 & 22.77 & 0.17 & 0.075 & 1 \\
    18  & 10.87149 & 41.57662 & 22.77 & 0.07 & 22.78 & 0.36 & 21.69 & 0.17 &     &     & 0.166 & 2 \\
    36  & 11.11893 & 41.49611 & 22.59 & 0.07 & 22.13 & 0.27 & 22.13 & 0.21 &     &     & 0.087 & 2 \\
    48  & 11.11262 & 41.57697 & 20.71 & 0.04 & 20.38 & 0.12 & 20.10 & 0.09 &     &     & 0.078 & 2 \\
    98  & 10.90495 & 41.17475 & 19.61 & 0.07 & 19.56 & 0.05 & 19.90 & 0.08 & 19.80 & 0.05 & 0.053 & 1 \\
    122 & 11.02539 & 41.21954 & 20.34 & 0.07 & 19.86 & 0.09 & 20.01 & 0.09 &     &     & 0.080 & 7 \\
    127 & 11.11795 & 41.32996 & 22.47 & 0.16 & 21.76 & 0.19 & 21.43 & 0.15 &     &     & 0.153 & 7 \\
    164 & 11.28475 & 41.66064 & 22.02 & 0.05 & 21.11 & 0.17 & 21.16 & 0.14 &     &     & 0.133 & 2 \\
    166 & 10.82593 & 41.18186 & 21.06 & 0.08 & 20.82 & 0.09 & 20.53 & 0.10 & 21.14 & 0.08 & 0.083 & 1 \\
    186 & 10.97685 & 41.25383 & 20.93 & 0.08 & 20.88 & 0.09 & 22.03 & 0.20 & 21.54 & 0.10 & 0.055 & 1 \\
    \enddata
    \tablecomments{a: photometry errors are given before inclusion of systematic error; b: Source ID is from \citet{Johnson}; sources measured in multiple fields have an "A" following their ID number. c: RA and DEC are measured in degrees.}
    \tablecomments{Table 1 is published in its entirety on-line in machine-readable format.
      A portion is shown here for guidance regarding its form and content.}
\end{deluxetable*}
\end{longrotatetable}

\begin{deluxetable*}{cccccccccccccccc}
\tablecaption{UVIT-PHAT stellar cluster properties.\label{tab:Properties}}
\tablewidth{700pt}
\tabletypesize{\scriptsize}
    \tablehead{ \colhead{Source ID$^a$} & \colhead{$\chi^2$} & \colhead{log(Z)} & \colhead{log(Z)$_{err}$} &  \colhead{log(age)} & \colhead{log(age)$_{err}$} & \colhead{log(M)} & \colhead{log(M)$_{err}$} &\colhead{E(B-V)} & \colhead{E(B-V)$_{err}$}  
    }
    \startdata
16 & 20.2 & 0.12 & 0.20 & 7.73 & 0.09 & 3.70 & 0.15 & 0.53 & 0.05 \\
18 & 76.1 & -0.19 & 0.21 & 6.96 & 0.42 & 2.68 & 0.27 & 0.51 & 0.13 \\
27 & 11.3 & 0.11 & 0.17 & 7.69 & 0.10 & 3.75 & 0.16 & 0.14 & 0.04 \\
36 & 104.7 & 0.18 & 0.10 & 8.29 & 0.10 & 3.05 & 0.09 & 0.00 & 0.09 \\
48 & 17.9 & 0.23 & 0.20 & 8.32 & 0.19 & 4.07 & 0.09 & 0.06 & 0.11 \\
93 & 1.6 & -1.35 & 0.38 & 8.12 & 0.28 & 4.22 & 0.11 & 0.34 & 0.10 \\
98 & 36.3 & 0.27 & 0.29 & 8.11 & 0.20 & 4.15 & 0.08 & 0.11 & 0.09 \\
127 & 1.0 & 0.13 & 0.33 & 8.40 & 0.29 & 3.78 & 0.15 & 0.19 & 0.15 \\
163 & 7.5 & -1.17 & 0.94 & 7.75 & 0.18 & 3.84 & 0.17 & 0.43 & 0.12 \\
164 & 16.8 & 0.25 & 0.21 & 8.50 & 0.16 & 4.00 & 0.11 & 0.03 & 0.21 \\
    \enddata
    \tablecomments{a: Source ID is from PHAT; sources measured in multiple fields have an "A" following their ID number.}
    \tablecomments{Table 2 is published in its entirety on-line in machine-readable format.
      A portion is shown here for guidance regarding its form and content.}
\end{deluxetable*}
 
\begin{deluxetable}{ccccccc}
\tablecaption{Statistics of Model Parameters for the UVIT-PHAT Cluster Fits\label{tab:FitStats}}
\tablewidth{700pt}
\tablehead{\colhead{Parameter} & \colhead{Mean} & \colhead{SD} \\ 
    }
    \startdata
$\chi^2$ &    9.8	&9.4&	\\
log(Z/Z$_{\odot}$) &	-0.14&	0.64 \\
log(Z/Z$_{\odot}$)$_{err}^{(a)}$&	0.24&	0.20\\
log(Age/yr)	&7.56	&0.77\\
Log(Age/yr)$_{err}^{(a)}$	&0.18	&0.10	\\
log(M/M$_{\odot}$)	&3.27&0.96\\
Log(M/M$_{\odot}$)$_{err}^{(a)}$	&0.15	&0.09\\
E(B-V)&	0.24	&0.15	\\
E(B-V)$_{err}^{(a)}$&	0.08&	0.04\\
    \enddata
    \tablecomments{$(a)$ The errors are denoted by subscript $_{err}$ and are 1$\sigma$ uncertainties.}
\end{deluxetable}

\end{document}